\documentclass[aps,pra,showpacs,english,10pt,a4,nofootinbib,notitlepage,twocolumn,superscriptaddress]{revtex4-1}
\usepackage[figuresright]{rotating}

\usepackage{graphicx}% Include figure files
\usepackage{dcolumn}% Align table columns on decimal point
\usepackage{bm}% bold math
\usepackage{hyperref}% add hypertext capabilities
\usepackage[mathlines]{lineno}% Enable numbering of text and display math
\usepackage{blkarray}
\usepackage{bbold}
\usepackage{epsfig}
\usepackage{epstopdf}
\usepackage{hhline}
\usepackage{makecell}
\usepackage{ulem}
\usepackage{array}
\newcolumntype{L}[1]{>{\raggedright\let\newline\\\arraybackslash\hspace{0pt}}m{#1}}
\newcolumntype{C}[1]{>{\centering\let\newline\\\arraybackslash\hspace{0pt}}m{#1}}
\newcolumntype{R}[1]{>{\raggedleft\let\newline\\\arraybackslash\hspace{0pt}}m{#1}}

\usepackage{bigdelim}
\newcolumntype{K}[1]{>{\centering\arraybackslash}p{#1}}
\newcolumntype{M}[1]{>{\centering\arraybackslash}m{#1}}

\usepackage[utf8]{inputenc}
\usepackage[T1]{fontenc}
\usepackage[english]{babel}  % english language

\usepackage{times}

\usepackage{amssymb,amsmath,amsfonts,amsthm,dsfont}

\usepackage{verbatim}
\usepackage{enumerate}

\graphicspath{{pics/}}
\usepackage{hyperref}
\usepackage{bbm}
\usepackage{booktabs}

\usepackage{ulem}

\usepackage[caption=false]{subfig}

\usepackage{color}
\definecolor{dred}{rgb}{.8,0.2,.2}
\definecolor{ddred}{rgb}{.8,0.5,.5}
\definecolor{dblue}{rgb}{.2,0.2,.8}
\definecolor{dgreen}{rgb}{.2,0.5,.2}
\newcommand{\be}{\begin{equation}}
\newcommand{\ee}{\end{equation}}

\begin{document}
\title{Gradient-based closed-loop quantum optimal control in a solid-state two-qubit system}
\date{\today}

\author{Guanru Feng} \thanks{Both authors contributed equally to this work.}
\affiliation{Institute for Quantum Computing, Waterloo, Ontario, N2L 3G1, Canada}
\affiliation{Department of Physics and Astronomy, University of Waterloo, Waterloo, Ontario, N2L 3G1, Canada}

\author{Franklin H. Cho} \thanks{Both authors contributed equally to this work.}
\affiliation{Institute for Quantum Computing, Waterloo, Ontario, N2L 3G1, Canada}
\affiliation{Department of Physics and Astronomy, University of Waterloo, Waterloo, Ontario, N2L 3G1, Canada}

\author{Hemant Katiyar}
\affiliation{Institute for Quantum Computing, Waterloo, Ontario, N2L 3G1, Canada}
\affiliation{Department of Physics and Astronomy, University of Waterloo, Waterloo, Ontario, N2L 3G1, Canada}

\author{Jun Li}
\affiliation{Department of Physics and Shenzhen Institute for Quantum Science and Engineering,
Southern University of Science and Technology, Shenzhen, 518055, China}
\affiliation{Institute for Quantum Computing, Waterloo, Ontario, N2L 3G1, Canada}
\affiliation{Department of Physics and Astronomy, University of Waterloo, Waterloo, Ontario, N2L 3G1, Canada}

\author{Dawei Lu}
\affiliation{Department of Physics and Shenzhen Institute for Quantum Science and Engineering,
Southern University of Science and Technology, Shenzhen, 518055, China}
\affiliation{Institute for Quantum Computing, Waterloo, Ontario, N2L 3G1, Canada}
\affiliation{Department of Physics and Astronomy, University of Waterloo, Waterloo, Ontario, N2L 3G1, Canada}

\author{Jonathan Baugh}
\email{baugh@uwaterloo.ca}
\affiliation{Institute for Quantum Computing, Waterloo, Ontario, N2L 3G1, Canada}
\affiliation{Department of Physics and Astronomy, University of Waterloo, Waterloo, Ontario, N2L 3G1, Canada}
\affiliation{Department of Chemistry, University of Waterloo, Waterloo, Ontario, N2L 3G1, Canada}

\author{Raymond Laflamme}
\email{laflamme@iqc.ca}
\affiliation{Institute for Quantum Computing, Waterloo, Ontario, N2L 3G1, Canada}
\affiliation{Department of Physics and Astronomy, University of Waterloo, Waterloo, Ontario, N2L 3G1, Canada}
\affiliation{Perimeter Institute for Theoretical Physics, Waterloo, Ontario, N2J 2W9, Canada}
\affiliation{Canadian Institute for Advanced Research, Toronto, Ontario M5G 1Z8, Canada}

\begin{abstract}
%Improving quantum control is very important to pave the way for scalable quantum information processing (QIP). 

Quantum optimal control can play a crucial role to realize a set of universal quantum logic gates with error rates below the threshold required for fault-tolerance. Open-loop quantum optimal control relies on accurate modeling of the quantum system under control, and does not scale efficiently with system size. These problems can be avoided in closed-loop quantum optimal control, which utilizes feedback from the system to improve control fidelity. In this paper, two gradient-based closed-loop quantum optimal control algorithms, the hybrid quantum-classical approach (HQCA) described in [Phys. Rev. Lett.~\textbf{118}, 150503 (2017)] and the finite-difference (FD) method, are experimentally investigated and compared to the open-loop quantum optimal control utilizing the gradient ascent method. We employ a solid-state ensemble of coupled electron-nuclear spins serving as a two-qubit system. Specific single-qubit and two-qubit state preparation gates are optimized using the closed-loop and open-loop methods. The experimental results demonstrate the implemented closed-loop quantum control outperforms the open-loop control in our system. Furthermore, simulations reveal that HQCA is more robust than the FD method to gradient noise which originates from measurement noise in this experimental setting. On the other hand, the FD method is more robust to control field distortions coming from non-ideal hardware.

\end{abstract}

\maketitle

Quantum computers are believed to outperform classical computers in solving certain problems~\cite{Algorithm}. However, turning theory into practice will require quantum processors that are resilient to noise. Fault tolerance theory assumes reasonable noise models and requires low error rates below a threshold level. Quantum optimal control~\cite{review,khaneja2005optimal,review2} is a useful tool to devise high fidelity control pulses that satisfy the threshold condition, and great progress has been made in different device architectures, e.g. superconducting qubits, quantum dots, ion traps, and nitrogen-vacancy centers~\cite{RBM_Martinis,silicon1,nv,ion}.

There are two broad classes of quantum optimal control: open-loop and closed-loop. Open-loop quantum control typically relies on accurate modelling of the system Hamiltonian and control parameters, therefore it may no longer produce expected result in realistic settings, e.g. effects due to the hardware tranfer function~\cite{Transfer1,Transfer2,PB}. In such cases, on one hand, open-loop optimization considering uncertainties in system and control Hamiltonians is being studied~\cite{MchL}; on the other hand, better performance can be achieved from closed-loop quantum control~\cite{Jun,review,Dawei,adaptive,close_superc}. Moreover, closed-loop quantum control combines the use of both classical and quantum resources in a way that the calculation remains efficient when the size of the system Hilbert space becomes classically intractable.
%>>>>>  not sure what you have in mind inthe previous sentence....
%>>>>>>
%The approach of combining classical and quantum resources was also proposed and implemented in quantum optimal control field . Current numerical approaches of quantum optimal control face scalability problems due to enormous computational resources required when the Hilbert space is large. With the quantum system under control fulfilling part of the optimization task, the classical resources needed is reduced significantly. Since the classical-quantum hybrid algorithms require feedback from quantum systems under control, they are close-loop control algorithms.

%\sout{The fundamental idea of quantum optimal control is to divide a pulse that accomplishes a gate in a sequence of small components and to modify these small components to achieve a predetermined measure of success in realizing the gate in question. The modification of these components can be done by searching the parameter space that characterizes the small components.  The methods to define the search direction are classified as gradient-based and non-gradient-based algorithms. In previous works~\cite{in-situ,Jun,NM1,NM2,NM3,WUREBING,ESR}, both the gradient-based and non-gradient-based algorithms were investigated in the closed-loop quantum control context.} 

In the context of closed-loop quantum optimal control, both gradient-based and gradient-free search algorithms have been investigated~\cite{in-situ,Jun,adaptive,close_superc,NM3,WUREBING,ESR}. Generally speaking, gradient-free algorithms converge slowly compared to gradient-based algorithms~\cite{Newgrad}. Gradient-based algorithms can be classified into two categories: model-free~\cite{in-situ} and model-dependent~\cite{Jun, WUREBING}. In their hybrid quantum-classical approach (HQCA), Li et al.~\cite{Jun} developed a scheme for measuring gradients based on the knowledge of the system Hamiltonian model, assuming a flat hardware transfer function. HQCA was successfully demonstrated in liquid-state nuclear magnetic resonance (NMR)~\cite{Jun,Dawei} where the transfer functions are relatively flat over the control frequency range, and improvement on control fidelity compared to open-loop control was observed. Ferrie et al.~\cite{in-situ} utilized a model-free algorithm stemming from the finite-difference (FD) method which uses finite differences to approximate derivatives/gradients. They also compared their FD method with gradient-free simplex or Nelder-Mead algorithm~\cite{adaptive,close_superc,NM3} and numerically showed that their method is more robust to control noise and requires fewer resources. % In Ref.~\cite{Dawei}, the control fidelity obtained using this method is higher than the open-loop optimal control method.

%\sout{a special pulse sequence to derive gradients where the pulse sequence depends on the system Hamiltonian model, assuming a flat hardware transfer function}. 

%It is straightforward to implement FI method in a close-loop quantum control by simply measuring the finite differences of the target function from the quantum system.
%However, it is still technically challenging to build large-scale qubit systems with long coherence time and good control abilities. The modest amount of quantum resources needed for solving practical problems, i.e. implementing practical quantum algorithms or quantum simulations,  is still beyond what can be reliably controlled nowadays. Recently, classical-quantum hybrid algorithms [], which release the requirement in the amount of qubits needed, attract a lot of interest. The basic idea is that advantages can be gained even with minimal quantum resources. 

In this paper, we experimentally investigated the two gradient-based closed-loop quantum control approaches, HQCA and FD, in a solid-state electron spin resonance (ESR) two-qubit system. The ensemble two-qubit system consists of hyperfine coupled electron and nuclear spins. This system combines advantages of electron spins and nuclear spins, i.e. large thermal polarization and fast control of electron spins, and long coherence time of nuclear spins. It has been shown that universal quantum control using only microwave excitation in this system is possible~\cite{hodges2008universal,zhang2011coherent,HBAC_daniel}. However, achieving high fidelity quantum control in a bulk ESR system is challenging. One reason is the limited frequency bandwidth of a conventional microwave resonator~\cite{PB}. Hence, when designing optimal control pulses for ESR systems, the hardware transfer function cannot usually be ignored. The HQCA method does not explicitly consider the transfer function, so it is important to test its performance experimentally and compare it with the FD method, which in principle accounts for the transfer function. We used two basis sets for the FD method, linear and Slepain~\cite{Slep1,Slep,Slep2}. The Slepain basis is designed for limited control bandwidth, and therefore can give further insights regarding the effects of the transfer function bandwidth. Finally, open-loop quantum optimal control using the first-order gradient ascent method was also implemented for comparison. In general, higher fidelities were observed with closed-loop control. Although the HQCA and FD methods showed similar experimental performance, we used simulations to find favorable conditions for each method. When the error in our system is dominated by the transfer function, the FD method performs better. When the error in our system is dominated by the randomly fluctuating measurement noise which results in errors in the measured gradients, the HQCA method performs better.

\textit{Two methods for deriving gradients---} The control problem we consider here is to prepare a desired state starting from a given initial quantum state. We choose the state fidelity defined in Eq.~(\ref{fidelity}) as the metric to evaluate the control quality,
\begin{align}
F = {\rm Tr}\left[U(T)\rho_iU(T)^{\dagger}\rho_f\right]/2^{n},
\label{fidelity}
\end{align} where $\rho_i$ and $\rho_f$ are the initial and target states, respectively, $T$ is the total duration of the control sequence, and $n$ is the number of qubits. $U(t)$ is the unitary evolution of the spin system in the presence of the system's internal Hamiltonian $H_0$ and control Hamiltonian $H_c\left({\bf u}(t)\right)$, and hence satisfies:
\begin{align}
\dot{U}(t) &= -i\left[H_0+H_c\left({\bf u}(t)\right)\right]U(t)\nonumber\\ U(0)&=I^{\otimes n}.
\label{evolution}
\end{align} Here ${\bf u}(t)$ is the collection of control parameters, e.g. control field amplitudes, and $I$ is the 2-dimensional identity operator. The goal is to maximize the fidelity defined in Eq.~(\ref{fidelity}). %\sout{Numerical methods are generally utilized for solving this problem as analytical solutions are very difficult to find in most of the cases. }

Gradient ascent pulse engineering (GRAPE)~\cite{khaneja2005optimal} is a well-known iterative numerical method to solve the optimization problem, where at the $q^{\operatorname{th}}$ iteration the control parameters are updated by
\begin{align}
{\bf u}^{q+1} &= {\bf u}^{q}+c_{q}{\bf g}^q,
\label{iterative}
\end{align}
where ${\bf g}^q$ is the gradient of $F$ with respect to the control parameters ${\bf u}^q$ and $c_{q}$ is an adaptive step size. Convergence happens at certain local optima and the solution can be accepted once the desired $F$ is realized with the parameters ${\bf u}^q$. In this paper, we refer to a class of numerical optimization methods which uses classical resources to calculate the fidelity $F$ and its gradients ${\bf g}^q$ as open-loop quantum optimal control.
 
One drawback of open-loop quantum optimal control is that it relies on accurate determination of $H_0$ and $H_c$, which can be difficult to obtain in real systems. Moreover, numerical methods become impractical when the size of the system is larger than a handful of qubits~\cite{Dawei}. To address such issues, Li et al. proposed a closed-loop quantum optimal control scheme known as HQCA which utilizes the quantum system under control as a quantum simulator in calculating the gradient ${\bf g}^q$. In the following, we briefly describe the method.

The HQCA approach can be applied to many quantum systems. Here, we take the spin-based magnetic resonance as an example. Consider a common control Hamiltonian in magnetic resonance systems, where the control magnetic field is in the transverse plane relative to the static magnetic field, i.e. in the $x$-$y$ plane:
\begin{align}
H_c(m)=\Sigma_{l=1}^N \left[u_{x,l}(m)\sigma_x^l+u_{y,l}(m)\sigma_y^l\right],
\end{align}
where $N$ is the number of spins that can be excited by the resonant alternating current (AC) magnetic field, $\sigma_\alpha^l$ is the Pauli operator of the $l^{\operatorname{th}}$ spin, and $u_{\alpha,l}(m)$ is the piecewise constant control amplitude ($\alpha=x$ or $y$). The unitary generated by the total Hamiltonian $H{=}H_0{+}H_c$ is then given by
\begin{align}
U(T)=\Pi_{m=1}^M e^{-i\Delta t\left[H_0+H_c(m)\right]},
\label{unitary}
\end{align}
where $\Delta t$ is the time step and $M$ is the total number of segments of $u_{\alpha,l}(m)$. The gradient at the $q^{\operatorname{th}}$ iteration, ${\bf g}^q$, is then defined as the partial derivative of $F$ with respect to $u_{\alpha,l}$:
\begin{align}
{\bf g}^q \equiv \left[\frac{\partial F}{\partial u_{x,l}(m)}, \frac{\partial F}{\partial u_{y,l}(m)}\right].
\label{HQCAg}
\end{align}
As proposed by Ref. \cite{Jun}, combining Eqs. (\ref{fidelity}) and (\ref{unitary}) gives
\begin{align}
\frac{\partial F}{\partial u_{\alpha,l}(m)}\approx\Delta t \frac{{\rm Tr}\left[\rho_{+\alpha}^l(m)\rho_f\right]-{\rm Tr}\left[\rho_{-\alpha}^l(m)\rho_f\right]}{2^n},\label{gradient}
\end{align}
where $\rho_{\pm\alpha}^l(m)=U_{m+1}^MR_\alpha^l(\pm \frac{\pi}{2})U_1^m\rho_i\left[U_{m+1}^MR_\alpha^l(\pm \frac{\pi}{2})U_1^m\right]^\dagger$.
This means that ${\bf g}^q$ can be experimentally measured by inserting $\pi/2$ rotations $R_\alpha^l(\frac{\pi}{2})$ and $R_\alpha^l(-\frac{\pi}{2})$ into the control pulse $U$. Since ${\bf g}^q$ is a $2NM$ dimensional vector, if we consider one experiment as containing the preparation of an initial state, implementation of the pulse and measurement over a chosen basis element, then at each iteration $4NMP$ experiments are required to obtain ${\bf g}^q$. Here, $P$ is the number of Pauli elements with non-zero coefficients that compose the target state $\rho_f$.

\begin{figure}[t]
%\centering
\includegraphics[width=0.48\textwidth]{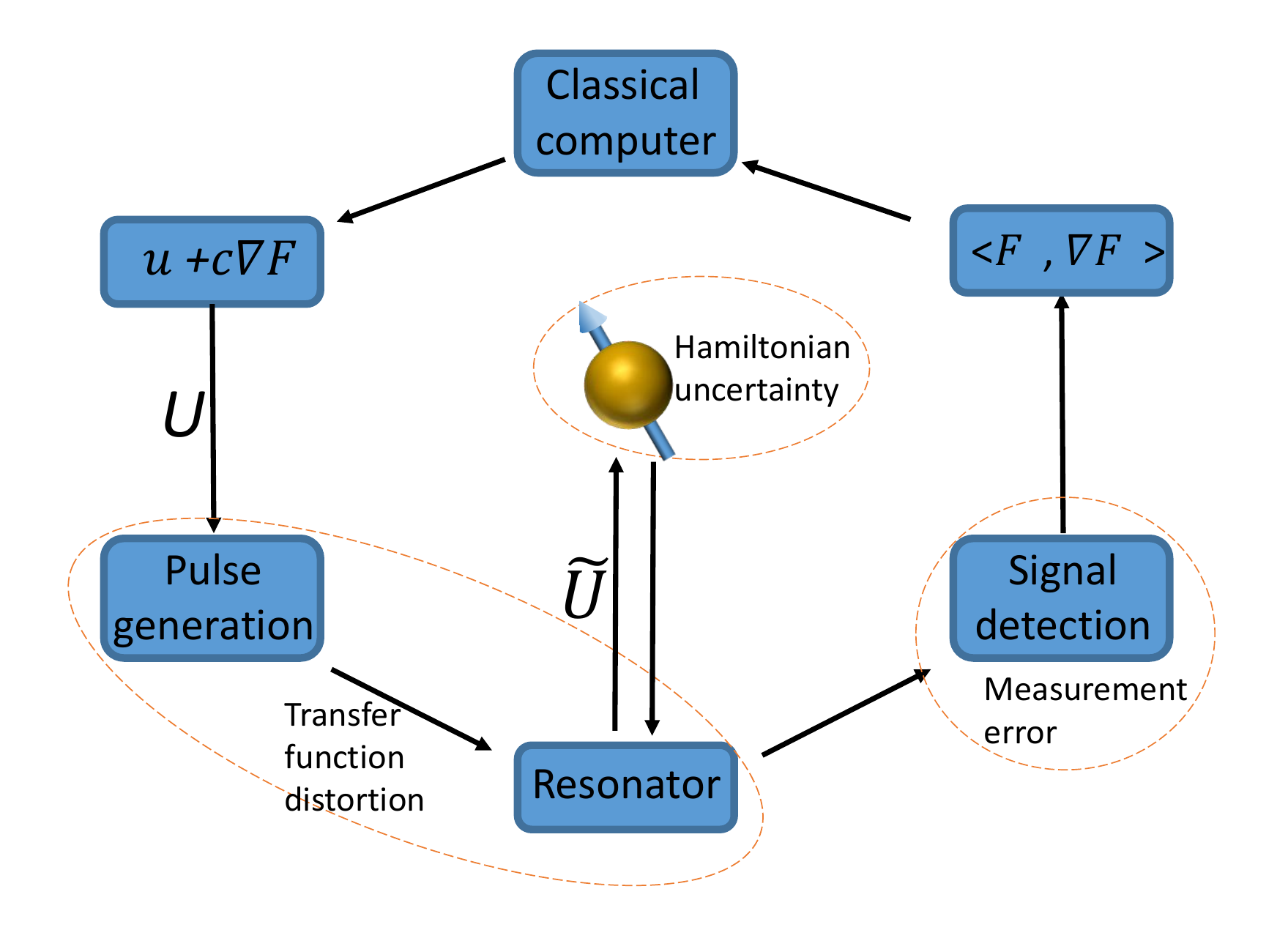}
\caption{(Color online) Flow diagram of gradient-based closed-loop optimal control as applied to ESR. Arrows label the direction of information flow. Error sources are labeled using orange dash-line circles. Both the HQCA and FD methods can compensate for the control errors caused by the uncertainty in the system Hamiltonian. Theoretically, the FD method finds the gradient $\partial F/\partial u$, while HQCA finds the gradient $\partial F/\partial \tilde{u}$. The pulse shape represented by $\tilde{u}$ is distorted by the hardware.} %The measurement error affects both methods but has a larger influence on the FD method.}
\label{closeloop}
\end{figure}

Figure~\ref{closeloop} shows the schematic of how closed-loop quantum optimal control is performed iteratively. While the HQCA method is a good choice for large systems with uncertain Hamiltonians~\cite{Jun,Dawei}, it does not account for hardware transfer function. In case the bandwidth of the hardware transfer function is much narrower than the intended bandwidth of the inserted $\pi/2$ rotation pulse, the scheme can fail. Even if the $\pi/2$ pulse functions properly, HQCA measures $\partial F/\partial \tilde{u}$ instead of $\partial F/\partial u$ (see Fig.~\ref{closeloop}), where $\tilde{u}$ represents the distorted pulse. Compared to the radio-frequency regime typical of NMR, the microwave transfer function relevant to ESR experiments tends to be much less uniform over the frequency range of interest. Thus, we consider another method of closed-loop control that can take these effects into account: the FD method. It uses finite differences to approximate the derivatives when an analytic expression of the gradient function is not available. If we write ${\bf u}^q{=}(u^q_{x,1},u^q_{y,1},\ldots, u^q_{x,N},u^q_{y,N})$ as the $q^{\operatorname{th}}$ control amplitude parameters, ${\bf g}^q$ can be expressed as
\begin{align}
{\bf g}^q &= \Sigma_kg_k^q{\bf v}^k,1{\leq}k{\leq}2NM \nonumber \\
g_k^q &\approx 
\frac{F({\bf u}^q+\Delta u_k^q {\bf v}^k) - F({\bf u}^q-\Delta u_k^q {\bf v}^k)}{2\Delta u_k^q}.
\label{finiteG}
\end{align}
Here $\{{\bf v}^k\}$ is a basis set that spans the parameter space and $\Delta u_k^q$ is a properly chosen difference value~\cite{in-situ}. To obtain the complete gradient vector ${\bf g}^q$, similar to the HQCA method, a total of $4NMP$ measurements are needed at each iteration. As the gradients are estimated directly from state fidelity measurements, distortions of the pulse due to the transfer function are accounted for in the process (see Appendix A). This method is useful when the hardware transfer function is strongly frequency dependent or is not accurately known.

%\sout{It should be mentioned that, different from the HQCA method, the derivation of the gradients from Eq.~(\ref{finiteG}) only depends on the $F$ values at different iterations, thus it is model-free and does not rely on any knowledge of how the target function depends on the parameters. Therefore, the distortion to the parameter space by the hardware does not affect the effectiveness of the derived gradients. This feature makes this method a good candidate for the gradient-based closed-loop optimal control in cases where the hardware transfer function cannot be measured with high accuracy.} \red{As the gradients are estimated from direct fidelity measurements, distortions of the pulse due to the transfer function is included in the process} -  \blue{\textbf{maybe need to rephrase/elaborate more..}}

In Ref.~\cite{in-situ}, the authors simulated a closed-loop (\textit{in-situ}) optimization scheme based on the FD method. Instead of using a complete basis set per iteration, they acquired gradient and performed optimization only with one random element of the basis set at a time. This results in fewer experiments per iteration, but convergence is slow if the random elements are not well chosen. In this work, we use two different basis sets: the first is a complete basis set in the time domain with a dimension of $2NM$ (we call it the linear basis), and the second is the Slepian basis~\cite{Slep1,Slep,Slep2}. The Slepian basis can be constructed to have fewer elements than $2NM$ with narrower control bandwidth, and is thus suitable for applications when the bandwidth is experimentally limited (see Appendix C). Roughly speaking, the Slepian basis set is a low-pass filtered version of the linear basis as the elements of the Slepian basis exhibit much smoother amplitude changes in the time domain compared to the linear basis.

%\textit{Two-qubit ESR system---}

\begin{figure*}[t]
%\centering
\includegraphics[width=0.98\textwidth]{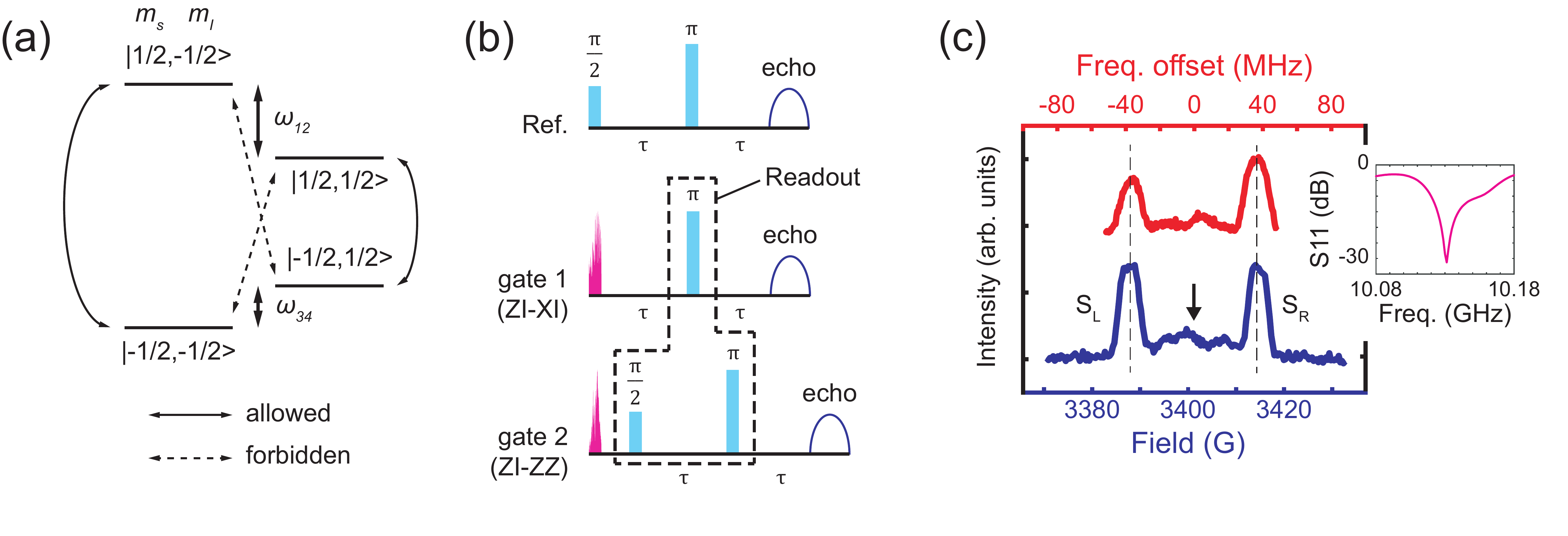}
\caption{(Color online) (a) Energy level diagram of the two-qubit system. Each level is designated by electron and nuclear spin quantum number, $m_s$ and $m_I$, respectively. (b) Pulse sequence used for the reference (spin echo), gate 1 ($\operatorname{ZI}$ to $\operatorname{XI}$), and gate 2 ($\operatorname{ZI}$ to $\operatorname{ZZ}$) measurement. For the reference and readout, 200 ns long square $\pi$/2 and $\pi$ pulses with excitation bandwidth of 5 MHz were used throughout the experiments, providing selective excitation of one allowed transition without affecting the other. The echo is formed after $\tau$=1 $\mu$s from the $\pi$ pulse. Phase cycling was implemented in gate 2 measurement to remove possible signal contributions from the transversal polarizations, which we do not want to measure. (c) ESR spectra acquired by sweeping the magnetic field (blue, lower trace) and frequency (red, upper trace) using the reference spin echo sequence. For the field swept spectrum, the microwave pulse frequency is fixed at the resonance frequency of the loop-gap resonator, determined from a separate microwave reflection (S11) measurement as shown in the inset. For the frequency swept spectrum, $B_0$ is fixed at 3401 G (denoted by the arrow in the figure). Two strong signals at 3401$\pm$14 G or $\pm$36 MHz ($S_L$ and $S_R$) correspond to the resonance condition of the two allowed ESR transitions. The unequal intensities obtained in the frequency swept spectrum are due to the frequency dependence of the spectrometer's transfer function.}
\label{fig1}
\end{figure*}

\textit{Experimental Results---} Experiments were carried out using a custom pulsed ESR spectrometer operating at X-band. An arbitrary waveform generation (AWG) enables pulse shaping, and a loop-gap resonator with Q$\sim$100 allows excitation over a bandwidth $\sim$100 MHz~\cite{ESRRB}. The sample we use is a single crystal of unlabeled malonic acid (CH$_2$(COOH)$_2$), where paramagnetic defects are created by gamma-ray irradiation~\cite{HBAC_daniel}. Since the carbon atoms are not spin labeled, all hyperfine couplings involve surrounding hydrogen atoms (I=1/2), and up to 8 have been observed~\cite{MAENDOR}. The general spin Hamiltonian can be written as \begin{align}
H_0 = \mu_B \bm{B_0}^{T} {\cdot} \bm{g} {\cdot} \bm{S} + \sum_{i=1}^8 \left( \bm{S}^{T}{\cdot}\bm{A}^{i} \cdot \bm{I}^i - \mu_n g_n \bm{B_0}^{T} {\cdot} \bm{I}^i \right),
\label{hamiltonian_general}
\end{align}
%\begin{align}
%\nonumber H_0 &= \omega_I I_z + A S_z I_z + B S_z I_x \\
%	&= \left[ \begin{array}{cccc}
%	\frac{\omega_I}{2}+\frac{A}{4} & \frac{B}{4} & & \\
%	\frac{B}{4} & -\frac{\omega_I}{2}+\frac{A}{4} & & \\
%	 & & \frac{\omega_I}{2}-\frac{A}{4} & -\frac{B}{4} \\
%	 & & -\frac{B}{4} & -\frac{\omega_I}{2}+\frac{A}{4}
%	\end{array} \right]
%	\label{hamiltonian}
%\end{align} 
where $\mu_B$ is the Bohr magneton, $\mu_n$ is the nuclear magneton, $\bm{B_0}{=}B_0\hat{z}$ is the externally applied magnetic field, $\bm{g}$ is the g-tensor of the electron spin, $g_n$ is the g-factor of the nuclear spin, $\bm{S}{=}(\hat{S}_x, \hat{S}_y, \hat{S}_z)$ is the electron spin operator, $\bm{A}^i$ and $\bm{I}^i$ are the hyperfine tensor and nuclear spin operator for the $i^{\operatorname{th}}$ nuclear spin, respectively (vectors are in bold). The hyperfine coupling to the $\alpha$-proton dominates, as it is about 10 times stronger than the second largest coupling. Therefore, we neglect all other protons and write a simplified, two-spin Hamiltonian: \begin{align}
H_0 = \omega_I I_z + A S_z I_z + B S_z I_x.
\label{hamiltonian_simplified}
\end{align} This Hamiltonian is written in the rotating frame of the electron and makes use of the secular approximation. Here $A$ and $B$ are the secular and pseudo-secular hyperfine couplings, respectively, and $\omega_I{=}\mu_ng_nB_0$ is the nuclear Zeeman frequency. At X-band where the strength of the static magnetic field ($B_0$) is around 0.34 T for $g{\sim}$2, $|\omega_I|{\sim}$14.5 MHz. Diagonalizing $H_0$ gives 
%\begin{align}
%\nonumber H_0^d &= \left[ \begin{array}{cccc}
%	\frac{\omega_{12}}{2} & & & \\
%	& -\frac{\omega_{12}}{2} & & \\
%	& & \frac{\omega_{34}}{2} & \\
%	& & & -\frac{\omega_{34}}{2}
%	\end{array} \right] \\
%	\nonumber |\omega_{12}| &=\sqrt{(\omega_I+A/2)^2+B^2/4} \\
%|\omega_{34}| &=\sqrt{(\omega_I-A/2)^2+B^2/4}.
%\label{hamiltonian_dia}
%\end{align}
\begin{align}
H_0^d &= \rm{Diag} \left[\frac{\omega_{12}}{2}, -\frac{\omega_{12}}{2}, \frac{\omega_{34}}{2}, -\frac{\omega_{34}}{2} \right] \nonumber \\ 
|\omega_{12}| &=\sqrt{(\omega_I+A/2)^2+B^2/4} \nonumber \\
|\omega_{34}| &=\sqrt{(\omega_I-A/2)^2+B^2/4},
\label{hamiltonian_dia}
\end{align} where the superscript `{\it d}' denotes the diagonal form. Figure~\ref{fig1}(a) shows the energy level diagram for the hyperfine coupled electron-nuclear spin system. The nuclear frequencies $\omega_{12}$ and $\omega_{34}$ are given in Eq.~(\ref{hamiltonian_dia}). The hyperfine coupling is known to be strongly anisotropic~\cite{ESRRB}, so the values of $A$ and $B$ depend on how the sample crystal is oriented with respect to $\bm{B_0}$. We chose an orientation in which $A{\gg}B$. Under such condition, it is more difficult to fully characterize the Hamiltonian experimentally compared to the cases when $A{\sim}B$. The uncertainty in the Hamiltonian provides a good testbed for comparing feedback control schemes. 

As shown in Fig.~\ref{fig1}(b), a spin echo sequence is utilized to read out the intensity of a particular ESR transition. When either the dc magnetic field, $B_0$, or the microwave frequency is varied to satisfy the resonance condition, strong peaks corresponding to the two allowed ESR transitions appear, as shown in Fig.~\ref{fig1}(c). In addition, smaller signals from the forbidden transitions appear between the two strong allowed peaks.

From spectral fitting, the forbidden transition rates are estimated to be $<$5\% of the allowed transition rates. The estimated range of $A$ and $B$ is 72${>}|A|{>}$66 MHz and 0${<}|B|{<}$26 MHz, where one constraint is that $|\omega_{12}|{+}|\omega_{34}|{=}$72 MHz (separation of the two allowed transitions). Since the forbidden transition rates are small, no electron spin echo envelope modulation (ESEEM) signals were observed. The lack of information from an ESEEM experiment is a key reason that the Hamiltonian parameters cannot be determined more accurately in this orientation.

\begin{figure}[t]
%\centering
\includegraphics[width=0.38\textwidth]{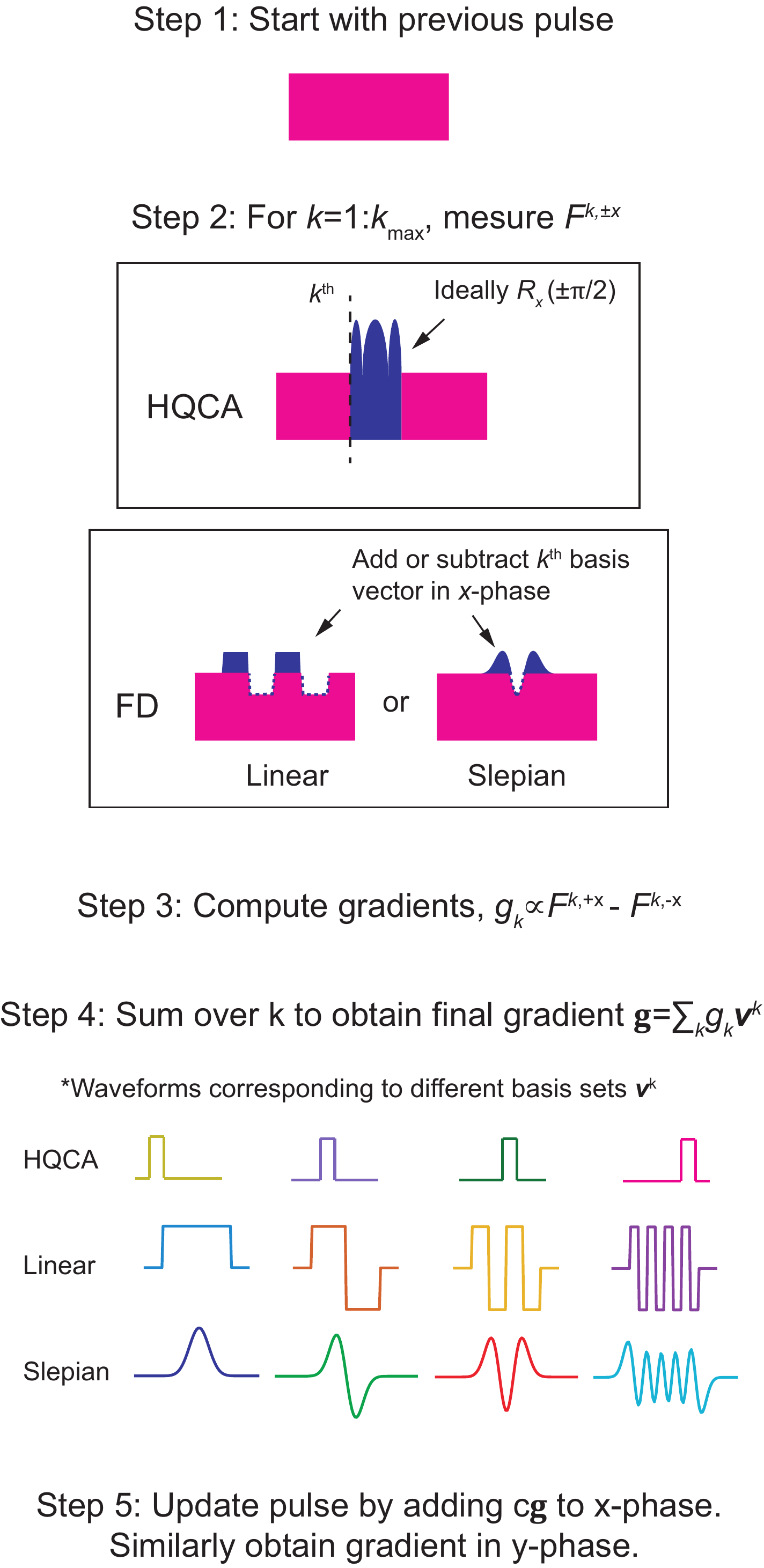}
\caption{(Color online) Process flow for optimizing closed-loop control. Starting with the pulse obtained from previous iteration (Step 1; shown here is a square pulse as an example), two control qualities $F^{k,\pm x}$ are measured (Step 2) to calculate the gradient in a particular direction of a basis denoted as ${\bf v}^{k}$ (Step 3). For the HQCA method, a $\pm \pi/2$ rotation pulse around the $x$-axis ($R_x(\pm \pi/2)$; see Appendix B) is inserted at the $k^{\operatorname{th}}$ segment. For the FD method, the pulse is perturbed by the addition/subtraction of a small amplitude vector proportional to the $k^{\operatorname{th}}$ basis element in the $x$-phase. Steps 2 and 3 are repeated for all elements of the basis, and the final gradient is obtained by summing all gradients (Step 4). The pulse is updated by adding the final gradient in $x$-phase (Step 5). A similar procedure is carried out to obtain the final gradient in $y$-phase. More details of the procedure, including information of different basis sets used, are given in the main text.}
\label{fig2}
\end{figure}

Finally, we test and compare open- and closed-loop quantum optimal control of two state-to-state gates on the two-qubit system described above. We denote the thermal equilibrium state by $\operatorname{ZI}$, where the first (second) letter refers to the state of electron (nuclear) spin. Here, we use the deviation density matrix to describe a state, and X, Y, and Z stand for the Pauli matrices. Gate 1 is the transformation $\operatorname{ZI}{\rightarrow}\operatorname{XI}$, and gate 2 is the transformation $\operatorname{ZI}{\rightarrow}\operatorname{ZZ}$. Both target states only contain one Pauli element and thus $P$=1. Characterizing the control quality requires measurement of the final state. This is done via two separate, selective readouts of the allowed ESR transitions, which we denote $S_{L}$ and $S_{R}$ (see Fig.~\ref{fig1}(b)). The selective readout is accomplished by fixing $B_0$ at the center of the spectrum and varying the pulse frequency using the AWG. To compensate for the transmission/receiver transfer function that is not flat, $S_{L}$ and $S_{R}$ must be properly normalized by their thermal reference signal intensity, $\overline{S}_L$ and $\overline{S}_R$. From Eq. (\ref{fidelity}) we can define the control quality for gates 1 and 2 as: \begin{align}
F_{\operatorname{XI}} &= \frac{1}{2} \left( S_{L}/\overline{S}_L+S_{R}/\overline{S}_R \right) \nonumber \\
F_{\operatorname{ZZ}} &= \frac{1}{2} \left( S_{L}/\overline{S}_L-S_{R}/\overline{S}_R \right).
\label{control_quality}
\end{align} The minus sign in $F_{\operatorname{ZZ}}$ is due to the fact that the two allowed transition peaks have opposite sign in the ideal spectrum for the state $\operatorname{ZZ}$. We note that while $F_{\operatorname{ZZ}}$ can be considered as the true state fidelity as defined in Eq.~(\ref{fidelity}), $F_{\operatorname{XI}}$ should be treated as a relative measure only. The reason is that $F_{\operatorname{XI}}$ can be larger than 1 when gate 1 performs better in exciting the transitions than the square $\pi/2$ pulse used in the reference measurement (see Fig.~\ref{fig1}(b)).

%One thing to note is that $F_{\operatorname{ZI-XI}}$ and $F_{\operatorname{ZI-ZZ}}$ are not "true" state fidelities, but relative measures of characterizing the performance of $\operatorname{ZI-XI}$ and $\operatorname{ZI-ZZ}$ gate. Moreover, direct comparison of $F_{\operatorname{ZI-XI}}$ and $F_{\operatorname{ZI-ZZ}}$ values may not be sought after because the readout scheme of the reference and $\operatorname{ZI-ZZ}$ spectrum is different from that of the $\operatorname{ZI-XI}$ spectrum (see Fig.~\ref{fig1}(b)).
% and it is not surprising there are numbers that is larger than 1 in Tab. \ref{Exp}

\begin{figure*}[t]
%\centering
\includegraphics[width=0.95\textwidth]{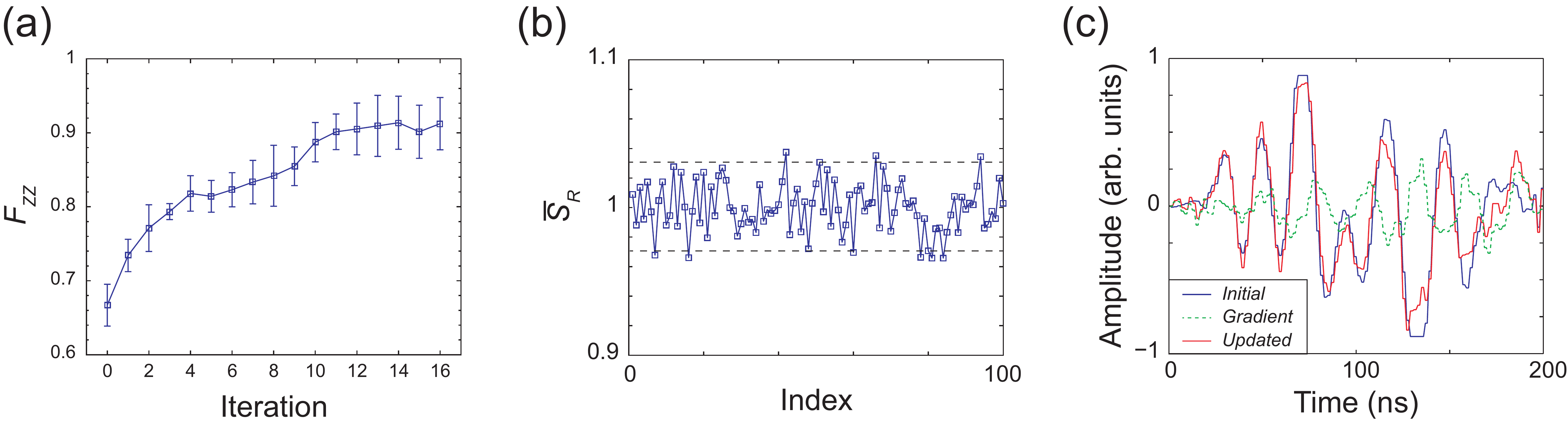}
\caption{(Color online) (a) $F_{\operatorname{ZZ}}$ as a function of the number of iterations (HQCA method). Here, the fidelity measurement for each pulse shape, together with the reference signal measurements, was repeated 5 times for the first 13 iterations and 50 times for the last 4 iterations. Each individual measurement was an average of 16,000 phase-cycled scans. The error bars indicate one standard deviation of the measurement results obtained over the repetitions (5 or 50). (b) Fluctuation of normalized reference signal intensity (here $\overline{S}_R$) during 100 repetitions. Dashed lines indicate $\pm0.03$ deviation from the normalized mean. (c) Example of pulse update. The blue curve represents the initial $x$-phase shape. The green dotted curve is the experimentally measured gradient $\bf g$. The red curve is the updated pulse obtained by adding $c\bf g$.}
\label{fig4}
\end{figure*}%Typically 8,000 averaging of phase-cycled scans was used per each repetition in measurements involving gradient calculations ($F_{\operatorname{ZI-XI}}^{ik,\pm x \operatorname{or} y}$ and $F_{\operatorname{ZI-ZZ}}^{ik,\pm x \operatorname{or} y}$), 16,000 averaging was used per each repetition in final measurements of updated $\operatorname{ZI-XI}$ and $\operatorname{ZI-XI}$ gate ($F^i_{\operatorname{ZI-XI}}$ and $F^i_{\operatorname{ZI-ZZ}}$).  

Figure~\ref{fig2} shows the general procedure of the closed-loop gate finding process. Below describes each step in detail. \newline \textbf{Step 1:} We start with the pulse obtained from the previous iteration. \newline \textbf{Step 2:} Next, we measure two control qualities $F^{k,\pm x}$. For HQCA, the pulse is obtained by inserting a $\pm\pi/2$ rotation around the $x$-axis at the $k^{\operatorname{th}}$ segment of the $x$-phase of the previous pulse (see Appendix B). For the $y$-phase pulse, a $\pm\pi/2$ rotation about the $y$-axis would be used. For the FD method, the the pulse is obtained by adding/subtracting a vector that is proportional to the $k^{\operatorname{th}}$ basis vector from the $x$ or $y$-phase of the previous pulse. The index $k$ runs from 1 to the total number of pulse segments for the HQCA method, or from 1 to the size of the basis set ${\bf v}^{k}$ for the FD method. \newline \textbf{Step 3:} The $x$-phase gradient is given by $g_k{\propto}F^{k,+x}{-}F^{k,-x}$ (see Eq.~(\ref{gradient}) for HQCA or Eq.~(\ref{finiteG}) for the FD method). \newline \textbf{Step 4:} By summing over all $k$, we obtain the full gradient for the current iteration in $x$-phase, ${\bf g}{=}\sum_{k}g_k{\bf v}^{k}$. For the HQCA method, ${\bf v}^{k}$ is a unit vector with the only non-zero element being 1 at the $k^{\operatorname{th}}$ index, e.g. $\{$0,0,1,0,...,0$\}$ for $k$=3. For the FD method, it is the $k^{\operatorname{th}}$ basis vector from a chosen set; in this paper we use linear and Slepian basis (see Appendix C for more detail). \newline \textbf{Step 5:} We update the $x$-phase pulse from previous iteration by adding $c{\bf g}$. Here, $c$ is a scaling factor chosen to avoid over- or undershooting. The $y$-phase pulse is updated in a similar manner, after finding the $y$-phase gradient.

%composite pulse Ideally this should realize the global $\pm\pi/2$ rotation, but because of the transfer function the observed control quality in real experiments is only around 0.5 \textcolor{red}{check this!})
% $B_0$ is fixed Relatively long square pulses of 100ns are used in the sequence to ensure narrow and selective excitation of the transitions, and the frequency of the pulses is altered appropriately between the separate measurement of the transitions while the magnetic field is fixed (see Fig.~1(c)). 

Figure~\ref{fig4} summarizes the closed-loop optimization of the $\operatorname{ZI}{\rightarrow}\operatorname{ZZ}$ pulse (gate 2). Similar to open-loop methods like GRAPE, the fidelity is seen to increase quickly in the first few iterations, but slows down and eventually saturates when the measured gradient becomes comparable to the shot noise. Although there is a convergence proof~\cite{in-situ, SPSA} in case of noisy measurements, there are two problems in practice: (1) inaccuracy in measured gradients and (2) difficulty in verifying small improvements in $F$. Moreover, a long-term drift in measurements can prevent $F$ from reaching a convergence~\cite{in-situ}. In practice, we found that there was no benefit in going beyond $\sim$15 iterations for the gates and protocols tested here, i.e. when the improvement in $F$ per iteration is smaller than the measurement noise.

Table~\ref{Exp} summarizes the final $F_{\operatorname{XI}}$ and $F_{\operatorname{ZZ}}$ obtained using three closed-loop optimal quantum control methods, (i) HQCA, (ii) FD with linear basis, and (iii) FD with Slepian basis. In addition, the results of open-loop control are presented. Here, the open-loop pulses were designed under three different conditions: (i) $A{=}$72 MHz, $B{=}$0 MHz, $\mathcal{T}{=}$1, (ii) $A{=}$66 MHz, $B{=}$26 MHz, $\mathcal{T}{=}$1, and (iii) $A{=}$66 MHz, $B{=}$26 MHz, $\mathcal{T}{=}\mathcal{T}_{meas}$. $\mathcal{T}$ denotes the transfer function, and $\mathcal{T}{=}$1 means that the pulses were designed under an ideal, flat transfer function. $\mathcal{T}{=}\mathcal{T}_{meas}$ indicates that the pulse design accounted for the experimentally measured transfer function (see Appendix E). In all the three conditions, a $\sim$10 MHz FWHM Larmor frequency distribution with the Lorentzian shape was considered. When designing the open-loop pulses, the gradients were calculated numerically using the FD method and two basis sets, the linear and Slepian basis, were used for comparison with their closed-loop counterparts. All the designed open-loop pulses have a simulated control quality higher than 1.020 for gate 1 and higher than 0.990 for gate 2 under their specific design conditions. As mentioned earlier section, the control quality of gate 1 is greater than 1 because it is more efficient in exciting the transitions than the square $\pi/2$ pulse used in the reference measurement (see Fig.~\ref{fig1}(b)).

First of all, we remark again that direct comparison of $F_{\operatorname{XI}}$ and $F_{\operatorname{ZZ}}$ is not proper as $F_{\operatorname{XI}}$ is not strictly a fidelity. For open-loop quantum optimal control, we observe higher $F$ when more accurate Hamiltonian parameters and realistic $\mathcal{T}$ are taken into consideration. However, closed-loop quantum optimal control methods still outperform the best open-loop results. Both closed-loop methods, HQCA and FD, produced similar control qualities under the experimental conditions tested here. However, simulations show that under different conditions, one method can perform better. This is described in the section below.

% , especially when the bandwidth-limited basis was used in design, which is expected as $\mathcal{T}_{meas}$ only corrects the pulse within its own limited bandwidth. 

%\sout{Not much difference among the three methods were observed, i.e., similar level of control quality could be reached with any of the three methods for a particular gate (within the standard deviations). As mentioned in the previous section, it is hard to directly compare $\operatorname{ZI-XI}$ and $\operatorname{ZI-ZZ}$ gate because the readout sequence of the two gates is not the same. }

%\begin{figure}[h!]
%\centering
%\includegraphics[width=0.5\textwidth]{exp_onequbit.png}
%\caption{The experimental results for one-qubit operation. Normalized signal strength is given for each case.}
%\label{fig2}
%\end{figure}

\textit{Simulations---} Simulations were performed to further elucidate the roles of measurement noise and the spectrometer transfer function in limiting the final closed-loop control quality. In these simulations, the closed-loop optimization is performed in the same way as before, but with the experimental system response simulated by computer. The simulations were ended when the overall improvement after five successive iterations is smaller than 0.01. As shown in Table~\ref{Simulation}, the simulation results indicate that HQCA is more robust to the measurement noise than FD methods. We find that this increased robustness for HQCA is due to its larger gradients compared to the FD methods.

\begin{table*}[t]
\normalsize
\centering
\begin{tabular}{C{9mm} | C{17mm} C{17mm} C{20mm} C{0.5mm} C{34mm} C{34mm} C{34mm}}
	\hline\hline
	 & \multicolumn{3}{c}{\textbf{Closed-loop control}} & \multicolumn{1}{c|}{} & \multicolumn{3}{c}{\textbf{Open-loop control}} \\
	 \cline{2-8}
	 & HQCA & FD (linear) & FD (Slepian) & \multicolumn{1}{c|}{} & $A{=}$72 MHz, \newline $B{=}$0 MHz, $\mathcal{T}{=}$1 & $A{=}$66 MHz, \newline $B{=}$26 MHz, $\mathcal{T}{=}$1 & $A{=}$66 MHz, \newline $B{=}$26 MHz, $\mathcal{T}{=}\mathcal{T}_{meas}$ \\
	 \hline
	 $F_{\operatorname{XI}}$ & 0.968(39) & 0.993(46) & 1.010(40) & \multicolumn{1}{c|}{} & $\cdot$ & 0.946(40)$^\dagger$\hspace{0.5mm}0.951(41)$^*$ & 0.955(37)$^\dagger$\hspace{0.5mm}0.990(51)$^*$ \\
	 $F_{\operatorname{ZZ}}$ & 0.914(36) & 0.918(43) & 0.932(37) & \multicolumn{1}{c|}{} & 0.807(37)$^\dagger$\hspace{0.5mm}0.799(37)$^*$ & 0.891(40)$^\dagger$\hspace{0.5mm}0.883(45)$^*$ & 0.889(39)$^\dagger$\hspace{0.5mm}0.902(43)$^*$\\
	 \hline\hline
	 
%	 & HQCA & FI & FI (Slepian) \\\hline
%closed loop & 0.9680(0.0391) & 0.9933(0.0462) & 1.0100(0.0400)  \\
%open loop, $\bf T$ & 0.9256(0.0371) & 0.9844(0.0369) & 0.9903(0.0512) \\
%open loop & 0.9416(0.0376) & 0.9501(0.0417) & 0.9512(0.0411) \\
%	\hline\hline
%	 & & & \\
%	 
%	
%	 \hline\hline
%	 & HQCA & FI & FI (Slepian) \\\hline
%closed loop & 0.9136(0.0359) & 0.9181(0.0428) &0.9323(0.0371)  \\
%open loop, $\bf T$  & 0.8820(0.0378) & 0.8961(0.0402) & 0.9017(0.0430) \\
%open loop & 0.9062(0.0437) & 0.8749(0.0366) & 0.8834(0.0445) \\
%open loop, no $B$ & 0.8126(0.0324) & 0.8006(0.0415) & 0.7990(0.0371) \\
%	\hline\hline
\end{tabular}
\caption{Experimental control qualities for gate operations 1 and 2. For closed-loop control, results are given for HQCA, FD with linear basis, and FD with Slepian basis. For open-loop control, the Hamiltonian parameters were varied as well as whether or not the spectrometer transfer function is accounted for in pulse design. The numbers reported here are averages of 50 measurements, and given in the parentheses are standard deviations (e.g. $0.968(39){=}0.968{\pm}0.039$). $^\dagger$ The open-loop control pulses were designed using the full-bandwidth basis set. $^*$ The open-loop control pulses were designed using the limited-bandwidth basis set, i.e. the Slepian basis, with a control bandwidth of 120 MHz. It should be noted that direct comparison of $F_{\operatorname{XI}}$ and $F_{\operatorname{ZZ}}$ is not proper as $F_{\operatorname{XI}}$ is not strictly a fidelity (see the main text).}
\label{Exp}
\end{table*}

The effect of the transfer function on closed-loop optimization was also tested in simulations. When the bandwidth of the transfer function is about twice the spectral width, both HQCA and FD methods give similar results. However, when the bandwidth of the transfer function becomes similar to the width of the spectrum, FD methods become superior to HQCA. This is because the derivation of Eq.~(\ref{gradient}) assumes $\mathcal{T}$=1, and non-ideal $\mathcal{T}$ will cause imperfections in $R_{\alpha}(\pm\pi/2)$ and measured gradients. In real experiments, if the transfer function is unknown or has limited accuracy, the distorted gradients cannot be corrected properly and will slow down the search. However, the gradient measurements should not be affected by non-ideal $\mathcal{T}$ in FD methods as these effects are accounted for automatically (see Appendix A).

Experimentally observed control qualities for gate 2 ($F_{\operatorname{ZZ}}$; see Eq.~(\ref{control_quality}) and Table~\ref{Exp}) are considerably lower than values obtained in the simulations. Transverse and longitudinal relaxations ($T_2{\sim}4$ $\mu$s and $T_1{\sim} 28$ $\mu$s) only give error of $\sim$2$\%$, which does not fully explain the discrepancy. This could be due to several reasons.For closed-loop methods, one reason is that long-term drifts in the spectrometer components during experiment (single iteration takes $\sim$8-12 hours) may introduce error. This limited the total number of iterations performed in experiment, which was far fewer compared to the simulations. The neglect of small couplings to nearby proton spins, i.e. the use of a simplified 2-spin Hamiltonian, can partly explain the imperfect control obtained with both open- and closed-loop methods (see Appendix I). Another possibility is that some components in the spectrometer transmission arm may exhibit small power non-linearities, so that $\mathcal{T}$ depends not only on frequency (as we assume) but also on microwave power.

%\hspace{0.1cm}

%Numerically, we compared/tested the two methods under different conditions/parameters. We found (a) Jun’s method is more robust to noise in the measurements, but the downside is that it requires a special pulse to obtain gradients, and is affected more by the bandwidth of the transfer function. (b) Finite difference method are more prone to noise. But the finite difference method, especially with bandwidth limited Slep basis set, can be friendlier/less demanding/easier to implement in a system with non-linear transfer function and/or limited control bandwidth (can also be less time-consuming for long pulses).

\textit{Conclusions---} Two gradient-based closed-loop quantum control methods are experimentally demonstrated in a solid-state two-qubit system and compared with gradient-based open-loop control methods. The implemented closed-loop control methods outperform the open-loop methods when the information of the Hamiltonian and hardware transfer function lacks accuracy in our quantum system. The open-loop control can be further improved by taking into consideration of the uncertainties of the transfer function and the Hamiltonian (e.g. distributions of $A$, $B$, and small couplings with the environment) which will increase the classical resources needed for the open-loop pulse design exponentially.

Together with simulations, we find that for the closed-loop optimzation, HQCA works better than FD methods when shot noise in measurements is large enough to be the dominant error limiting the gradient search. When $\mathcal{T}$ is narrow and/or the control bandwidth is limited (often by hardware), FD methods can perform better than HQCA. With the bandwidth-limited Slepian basis set, the gradient finding procedure of the FD method can be made less time-consuming and the pulses generated may be friendlier to implement in experiment. Thus, starting with a viable open-loop quantum optimal control (e.g. GRAPE) pulse and running subsequent iterations of an appropriate closed-loop control protocol may be a good strategy to reach high control quality under realistic experimental conditions~\cite{adaptive}. It should be mentioned that in the experimental setting of this work, the signal-to-noise ratio (SNR) is one of the major factors limiting the final control quality of the closed-loop methods (SNR$\sim$17 with 16,000 averaging). This also causes the closed-loop optimization processes to be more time-consuming than the open-loop counterparts. However, as mentioned earlier, our open-loop pulse design did not consider all of the factors to expediate the process. Moreover, the time consumption of open-loop control optimization process will increase exponentially with the system size, which is not the case with the closed-loop control process. Therefore, the time consumption is not a critical drawback of the closed-loop control. In the future, it may be of interest to combine the optimal random orientation method~\cite{SPSA,optimal SPSA} with the closed-loop quantum control methods for better efficiency of convergence rate.

\begin{table}[t]
\normalsize
\centering
\begin{tabular}{C{18mm} C{19mm} C{19mm} C{19mm}}
	\hline\hline
	Noise & HQCA & FD (linear) & FD (Slepian) \\\hline
${\sim}$0.03 & 0.958(02) & 0.967(01) &0.973(02)  \\
${\sim}$0.07 & 0.958(02) & 0.968(04) &0.973(03)  \\
${\sim}$0.14 & 0.957(03) & 0.947(26) &0.960(30) \\
${\sim}$0.20 & 0.956(09) & 0.926(48) &0.905(29) \\
	\hline\hline
	\\
	\hline\hline
	FWHM & HQCA & FD (linear) & FD (Slepian) \\\hline
$\sim$130MHz & 0.958 & 0.970 & 0.975 \\
$\sim$70MHz & 0.936 & 0.974 &0.964 \\
	\hline\hline
\end{tabular}
\caption{Simulation results showing the effects of noise level and transfer function bandwidth on the control fidelities $F_{\operatorname{ZZ}}$ for gate 2. Noise: Zero-mean Gaussian random noise with a standard deviation ($\sigma$) of 0.03, 0.07, 0.14, and 0.20 was added to the ideal gradient measurements, where the reference measurements ($\overline{S}_L$ and $\overline{S}_R$) are normalized to 1. The averages and standard deviations of ten trials in each condition are given. Transfer function: In each case, the entire control pulse sequence was distorted according to a realistic transfer function with full width at half maximum (FWHM) of 130 and 70 MHz. In the simulation with different noise levels, the transfer function with 130 MHz FWHM was considered. Simulated pulse finding was stopped when the overall improvement over five successive iterations was smaller than 1$\%$. $T_2$ was not included in this simulation.}
\label{Simulation}
\end{table}

%A great number of scans were taken for each measurment and hence the closed-loop optimization processes were time-consuming and the system drifts began to play a role. By enhancing the signal-to-noise ratio (e.g. working at low temperature), the closed-loop optimization can be comparable to or better than the open-loop optimization in time consumption. 
 
%\begin{table}[b]
%\normalsize
%\centering
%\begin{tabular}{C{18mm} C{18mm} C{18mm} C{19mm}}
%	\hline\hline
%	FWHM & HQCA & FD (linear) & FD (Slepian) \\\hline
%$\sim$130MHz & 0.958 & 0.975 &0.970  \\
%$\sim$70MHz & 0.936 & 0.964 &0.974 \\
%	\hline\hline
%\end{tabular}
%\caption{The simulation results of how the control quality of the different closed-loop quantum optimal control methods are affected by different transfer functions. The whole pulse sequences (including the closed loop control pulses, inserted $R^e_{x \operatorname{or} y}(\pm\pi/2)$ in the HQCA method, added/subtracted basis for the FD methods, and square pulses in the readout) were distorted according to realistically designed transfer function with full width at half maximum (FWHM) of approximately 130 and 70 MHz. Same criteria, i.e., the overall improvement after five successive iterations is smaller than 0.01, was used to determine when to stop the simulations. Values of $F_1$ are given only. $T_2$ was not considered in this simulation.}
%\label{Simulation:Transfer}
%\end{table}

{\bf Acknowledgement} This research was supported by NSERC, the Canada 
Foundation for Innovation, CIFAR, the province of Ontario, Industry Canada.

%\bibliography{UDA}

\begin{thebibliography}{10}
%%\bibitem{linearEq} A. W. Harrow, A. Hassidim, and S. Lloyd, Quantum Algorithm for Linear Systems of Equations, Phys. Rev. Lett. 103, 150502 (2009).  
\bibitem{Algorithm}
A.~Montanaro, npj Quantum Information~\textbf{2}, 15023 (2016).

\bibitem{review}
C.~Brif, R.~Chakrabarti, and H.~Rabitz, New J. Phys.~\textbf{12}, 075008 (2010).

\bibitem{review2} S. J. Glaser et al., Eur. Phys. J. D 69, 279 (2015).
\bibitem{khaneja2005optimal}
N.~Khaneja, T.~Reiss, C.~Kehlet, T.~Schulte-Herbr{\"u}ggen, and S.~J.~Glaser, J. Magn. Reson.~\textbf{172}, 296 (2005).

\bibitem{RBM_Martinis}
R.~Barends et al., Nature~\textbf{508}, 500 (2014).

\bibitem{silicon1}
M.~Veldhorst et al., Nature Nano.~\textbf{9}, 981(2014).
  
\bibitem{nv}
X.~Rong, J.~Geng, F.~Shi, Y.~Liu, K.~Xu, W.~Ma, F.~Kong, Z.~Jiang, Y.~Wu, J.~Du, Nature Communications~\textbf{6}, 8748 (2015).

\bibitem{ion}
A.~Bermudez et al., Phys. Rev. X~\textbf{7}, 041061 (2017).

\bibitem{Transfer1}
T.~Kaufmann, T.~J.~Keller, J.~M.~Franck, R.~P.~Barnes, S.~J.~Glaser, J.~M.~Martinis, and S.~Han, J. Magn. Reson.~\textbf{235}, 95 (2013).

\bibitem{Transfer2}
S.~Gustavsson, O.~Zwier, J.~Bylander, F.~Yan, F.~Yoshihara, Y.~Nakamura, T.~P.~Orlando, and W.~D.~Oliver, Phys. Rev. Lett.~\textbf{110}, 040502 (2013).

\bibitem{PB}
G.~Feng, J.~J.~Wallman, B.~Buonacorsi, F.~H.~Cho, D.~K.~Park, T.~Xin, D.~Lu, J.~Baugh, and R.~Laflamme, Phys. Rev. Lett.~\textbf{117}, 260501 (2016).

\bibitem{MchL} M. Y. Niu, S. Boixo, V. Smelyanskiy, and H. Neven, arXiv preprint arXiv:1803.01857 (2018).

\bibitem{Jun}
J.~Li, X.~Yang, X.~Peng, and C.-P.~Sun., Phys. Rev. Lett.~\textbf{118}, 150503 (2017).

\bibitem{adaptive}
D.~J.~Egger and F.~K.~Wilhelm, Phys. Rev. Lett.~\textbf{112}, 240503 (2014).

\bibitem{close_superc}
J. Kelly et al., Phys. Rev. Lett.~\textbf{112}, 240504 (2014).

\bibitem{Dawei}
D.~Lu et al., npj Quantum Information~\textbf{3}, 45 (2017).

\bibitem{Newgrad} Sh. Machnes, E. Ass$\acute{e}$mat, D. Tannor, and F. K. Wilhelm, Phys. Rev. Lett. 120, 150401 (2018).

\bibitem{ESR}
D.~L.~Goodwin, W.~K.~Myers, C.~R.~Timmel, and I.~Kuprov, arXiv:1707.05184v1.

%\bibitem{NM1}
%D.~J.~Egger and F.~K.~Wilhelm, Phys. Rev. Lett.~\textbf{112}, 240503 (2014).

%\bibitem{NM2}
%J.~Kelly et al., Phys. Rev. Lett.~\textbf{112}, 240504 (2014).

\bibitem{NM3}
M.~A.~Rol, et al., Phys. Rev. Applied~\textbf{7}, 041001 (2017).

\bibitem{WUREBING}
R.-B.~Wu, B.~Chu, D.~H.~Owens, and H.~Rabitz, arXiv:1712.01780.

\bibitem{in-situ}
C.~Ferrie and O.~Moussa, Phys. Rev. A~\textbf{91}, 052306 (2015).

\bibitem{hodges2008universal}
J.~S.~Hodges, J.~C.~Yang, C.~Ramanathan, and D.~G.~Cory, Phys. Rev. A~\textbf{78}, 010303(R) (2008).

\bibitem{zhang2011coherent}
Y.~Zhang, C.~A.~Ryan, R.~Laflamme, and J.~Baugh, Phys. Rev. Lett.~\textbf{107}, 170503 (2011).

\bibitem{HBAC_daniel}
D.~K.~Park, G.~Feng, R.~Rahimi, S.~Labruy\`{e}re, T.~Shibata, S.~Nakazawa, K.~Sato, T.~Takui, R.~Laflamme, and J.~Baugh, Quantum Information Processing \textbf{14}, 2435 (2015).
\bibitem{Slep1} D. Slepian, The Bell System Technical Journal 57, 1371 (1978).
\bibitem{Slep2} D. Slepian, SIAM Review 25, 379 (1983).
\bibitem{Slep}
D.~G.~Lucarelli, arXiv:1611.00188.
%\bibitem{Slep2} V.M. Frey, S. Mavadia, L.M. Norris, W. de Ferranti, D. Lucarelli, L. Viola, and M.J. Biercuk, Nature Communicationsvolume 8, 2189 (2017).


\bibitem{ESRRB}
D.~K.~Park, G.~Feng, R.~Rahimi, J.~Baugh, and R.~Laflamme, J. Magn. Reson.~\textbf{267}, 68 (2016).

%\bibitem{knill1997theory}
%E.~Knill, and R.~Laflamme, Phys. Rev. A 55~(2), 900 (1997).
\bibitem{MAENDOR}
E.~Sagstuen, A.~Lund, Y.~Itagaki, and J.~Maruani, J. Phys. Chem. A~\textbf{104}, 6362 (2000).
\bibitem{SPSA}
J.~C.~Spall, IEEE Trans. Autom. Control~\textbf{37}, 332 (1992).

\bibitem{optimal SPSA}
P.~Sadegh and J.~C.~Spall, IEEE Trans. Autom. Control~\textbf{43}, 1480 (1996).

%\bibitem{RBT2} B. R Johnson, M. P. da Silva, C. A. Ryan, S. Kimmel, J. M. Chow, and T. A. Ohki, New J. Phys. 17, 113019 (2015).

%\bibitem{twirl_science} J. Emerson, M. Silva, O. Moussa, C. Ryan, M. Laforest, J. Baugh, D. G. Cory, and R. Laflamme, Science 317 (5846), 1893-1896 (2007).

%\bibitem{twirl_solid} O. Moussa, M. P. da Silva, C. A. Ryan, and R. Laflamme, Phys. Rev. Lett. 109 (7), 070504 (2012).

%\bibitem{twirl_liq} D. Lu \textit{et al.}, Phys. Rev. Lett. 114 (14), 140505 (2015).

\expandafter\ifx\csname url\endcsname\relax
  \def\url#1{\texttt{#1}}\fi
\expandafter\ifx\csname urlprefix\endcsname\relax\def\urlprefix{URL }\fi
\expandafter\ifx\csname href\endcsname\relax
  \def\href#1#2{#2} \def\path#1{#1}\fi

%\bibitem{TroyPRL} T.W. Borneman,  C. E. Granade, and D. G. Cory, Parallel
%Information Transfer in a Multinode Quantum Information Processor, Physical
%Review Letters 108~(14) (2012) 140502.

%\bibitem{TroyThesis}
%T.~Borneman, Techniques for noise suppression and robust control in spin-based
%  quantum information processors, Ph.D. thesis, Massachusetts Institute of
%  Technology (2012).

%\bibitem{CPMGPRL}
%J. J. L. Morton, A. M. Tyryshkin, A. Ardavan, K. Porfyrakis, S. A. Lyon, G. A. D. Briggs, High Fidelity Single Qubit Operations Using Pulsed Electron Paramagnetic Resonance, Phys. %Rev. Lett. 95 (2005) 200501.

%\bibitem{RBM_Biercuk2009}
%M.~J. Biercuk, H.~Uys, A.~P. Vandevender, N.~Shiga, W.~M. Itano, J.~J.
%  Bollinger, High-fidelity quantum control using ion crystals in a penning
%  trap, Quantum Info. Comput. 9~(11) (2009) 920--949.

%\bibitem{PhysRevA.84.030303}
%K.~R. Brown, A.~C. Wilson, Y.~Colombe, C.~Ospelkaus, A.~M. Meier, E.~Knill,
%  D.~Leibfried, D.~J. Wineland, Single-qubit-gate error below $10^{-4}$ in a
%  trapped ion, Phys. Rev. A 84 (2011) 030303.

%\bibitem{negrevergne2006benchmarking}
%C.~Negrevergne, T.~S. Mahesh, C.~A. Ryan, M.~Ditty, F.~Cyr-Racine, W.~Power,
%  N.~Boulant, T.~Havel, D.~G. Cory, R.~Laflamme, Benchmarking quantum control
%  methods on a 12-qubit system, Physical Review Letters 96~(17) (2006) 170501.

%\bibitem{RBM_Schoelkopf}
%J.~M. Chow, J.~M. Gambetta, L.~Tornberg, J.~Koch, L.~S. Bishop, A.~A. Houck,
%  B.~R. Johnson, L.~Frunzio, S.~M. Girvin, R.~J. Schoelkopf, Randomized
%  benchmarking and process tomography for gate errors in a solid-state qubit,
%  Phys. Rev. Lett. 102 (2009) 090502.

%\bibitem{RBM_Martinis}
%R.~Barends, J.~Kelly, A.~Megrant, A.~Veitia, D.~Sank, E.~Jeffrey, T.~C. White,
%  J.~Mutus, A.~G. Fowler, B.~Campbell, Y.~Chen, Z.~Chen, B.~Chiaro,
%  A.~Dunsworth, C.~Neill, P.~O'Malley, P.~Roushan, A.~Vainsencher, J.~Wenner,
%  A.~N. Korotkov, A.~N. Cleland, J.~M. Martinis, Superconducting quantum
%  circuits at the surface code threshold for fault tolerance, Nature 508~(7497)
%  (2014) 500--503.

%\bibitem{RBM_opticallattice}
%S.~Olmschenk, R.~R~Chicireanu, K.~D. Nelson, J.~Porto, Randomized benchmarking
%  of atomic qubits in an optical lattice, New Journal of Physics 12~(11) (2010)
%  113007.

%\bibitem{silicon1}
%M. Veldhorst, J. C. C. Hwang, C. H. Yang, A. W. Leenstra, B. de Ronde, J. P. Dehollain,
%J. T. Muhonen, F. E. Hudson, K. M. Itoh, A. Morello, A. S. Dzurak, An addressable quantum dot qubit with fault-tolerant control-fidelity, Nature Nano.
%  9 (2014) 981-985.

%\bibitem{RBM_silicon}
%J.~T. Muhonen, A.~Laucht, S.~Simmons, J.~P. Dehollain, R.~Kalra, F.~E. Hudson,
%  S.~Freer, K.~M. Itoh, D.~N. Jamieson, J.~C. McCallum, A.~S. Dzurak,
%  A.~Morello, Quantifying the quantum gate fidelity of single-atom spin qubits
%  in silicon by randomized benchmarking, Journal of Physics: Condensed Matter
%  27~(15) (2015) 154205.

%\bibitem{nv}
%X. Rong, J Geng, F. Shi, Y. Liu, K. Xu, W. Ma, F. Kong,
%Z. Jiang, Y. Wu, J. Du, Experimental fault-tolerant universal quantum gates with solid-state spins under ambient conditions, Nature Communications 6 (2015) 8748.


\end{thebibliography}

\vskip 12pt

\clearpage

\onecolumngrid

\newcommand{\beginsupplement}{%
        \setcounter{table}{0}
        \renewcommand{\thetable}{S\arabic{table}}%
        \setcounter{figure}{0}
        \renewcommand{\thefigure}{S\arabic{figure}}%
     }

\beginsupplement

\begin{widetext}
\center
{\bf Appendix: Closed-loop quantum optimal control in a solid-state two-qubit system}
\medskip
\bigskip
\end{widetext}

\twocolumngrid
\appendix
\section{Difference between the gradient derived with the HQCA and FD methods}
As mentioned in the main text, HQCA does not consider the existence of hardware transfer function. In case of an extremely narrow transfer function, the required $\pm\pi/2$ pulses can be seriously distorted and the method can totally fail. When the distortion is mild, the $\pm\pi/2$ pulses can still function well, which is the case for our experiment. Here we describe the difference between the gradient derived with the HQCA and FD methods.

The target function, which is the state fidelity $F$, is a function of the control amplitudes $\bf{u}$, i.e. $F(\bf{u})$. In presence of hardware transfer function $\mathcal{T}$, it can be further written as $F(\tilde{\bf{u}}){=}F(\mathcal{T}(\bf{u}))$ where $\tilde{\bf{u}}$ is the distorted control amplitudes. In HQCA, the derived gradient is
\begin{align}
\tilde{g}=\frac{\partial F}{\partial \tilde{u}}.\label{HQCAderivative}
\end{align}
In the FD method, the derived gradient is
\begin{align}
g=\frac{\partial F}{\partial u}.\label{FDderivative}
\end{align}
The relationship between $\tilde{g}$ and $g$ is
\begin{align}
g=\tilde{g}\cdot\frac{\partial \tilde{u}}{\partial u},\label{grelation}
\end{align} where $\frac{\partial \tilde{u}}{\partial u}$ is given by $\mathcal{T}$. The gradient $g$ is the “correct” gradient for updating the control amplitudes $\bf{u}$. As shown with the simulation in the main text, narrower bandwidth of $\mathcal{T}$ yields lower final fidelity with HQCA while fidelities obtained using the FD method are practically independent of $\mathcal{T}$. This indicates that the gradient derived with the FD method automatically accounts for $\mathcal{T}$. If $\mathcal{T}$ is known accurately, obtaining $g$ from $\tilde{g}$ is possible using Eq.~(\ref{grelation}). 

%Therefore, we conclude that FD method accounts for the transfer function automatically while HQCA does not. If the information of $\mathcal{T}$ is limited in accuracy, FD method derives more accurate gradient.

\section{Non-ideal $\pm\pi/2$ rotation pulse in HQCA}
The pulse we insert in HQCA should ideally be a non-selective $\pm\pi/2$ rotation, but this is difficult to realize with simple square or Gaussian pulses due to the limited control bandwidth and pulse power. So we try to realize the rotation by simply adding two selective $\pm\pi/2$ square pulses at the two different frequencies, given by the distance between the two allowed ESR peaks from the center (i.e. $\pm$36 MHz; see Fig.~\ref{fig1}(c) of the main text). Even though the composed pulse does not provide accurate $\pm\pi/2$ rotation, the simulation and experiment confirm that it is sufficient in deriving gradient toward higher control quality. Here, we give a short proof for the effectiveness of a non-ideal $\pm\pi/2$ rotation in the HQCA method. From Ref.~\cite{Jun}, we have
\begin{align}
\frac{\partial F}{\partial u_{\alpha,l}(m)} & = \nonumber \\
& \frac{{\rm Tr}(-i \Delta t U_{m+1}^M[\sigma_\alpha^l,U_1^m\rho_iU_1^{m\dagger}]U_{m+1}^{M\dagger}\rho_f)}{2^n},\label{derivative}
\end{align}
and for any operator $\rho$,
\begin{align}
[\sigma_\alpha^l,\rho]=i[R_\alpha^l(\frac{\pi}{2})\rho R_\alpha^l(\frac{\pi}{2})^\dagger-R_\alpha^l(-\frac{\pi}{2})\rho R_\alpha^l(-\frac{\pi}{2})^\dagger].\label{comutor}
\end{align}
By combining Eqs.~(\ref{derivative}) and (\ref{comutor}), we obtain Eq.~(\ref{gradient}) in the main text. When the rotation angle $\theta{\neq}\pi/2$, it is easy to see that
\begin{align}
[\sigma_\alpha^l,\rho] = \frac{i[R_\alpha^l(\theta)\rho R_\alpha^l(\theta)^\dagger -R_\alpha^l(-\theta)\rho R_\alpha^l(-\theta)^\dagger]}{\sin(\theta)},
\end{align} meaning that the orientation of the derived gradient is unaffected (the amplitude is scaled by a factor of $\sin(\theta)$).

\section{Bandwidth-limited Slepian basis} 
The Slepian basis set we use in this work is composed of Slepian sequances which are also called discrete prolate spheroidal sequences~\cite{Slep1}. Slepian sequences with a sequence length $N$ and half bandwidth W$\in (0,0.5]$ are defined to be the real solutions to the eigenvalue problem
\begin{align}
\sum_{m=0}^{N-1} \frac{\sin 2\pi W(l-m)}{\pi (l-m)} & v_k(m;N,W) \nonumber \\ 
& =\lambda_k(N,W)v_k(l;N,W).\label{slepEq}
\end{align}
Here, $k,l{\in}\{0,1,...,N{-}1\}$, and $v_k(l;N,W)$ is the $l^{\operatorname{th}}$ element of the $k^{\operatorname{th}}$ order Slepian sequence. The Slepian sequences are spectrally concentrated in the frequency range $[-W/ \Delta t, W/ \Delta t]$, especially the first 2$NW$ ones. Ref.~\cite{Slep2} used the first 2$NW$ Slepian sequences to approximate the space of bandwidth limited sequences of length $N$.

\section{Linear basis} 
The shaped pulse used in this work has 100 segments with 2 ns step size. In addition to the badnwidth-limited Slepian basis described above, a linear basis set of dimension 100 based on the Hadamard basis is used. The complete space of this linear basis set is composed of three subspaces with 64-dimensional, 32-dimensional and 4-dimensional Hadamard basis spaces. The linear basis set is chosen this way because having as many non-zero elements as possible in the basis set is preferred to generate observable difference in the FD method.

\section{Transfer function} 

\begin{figure}[h!]
%\centering
\includegraphics[width=0.48\textwidth]{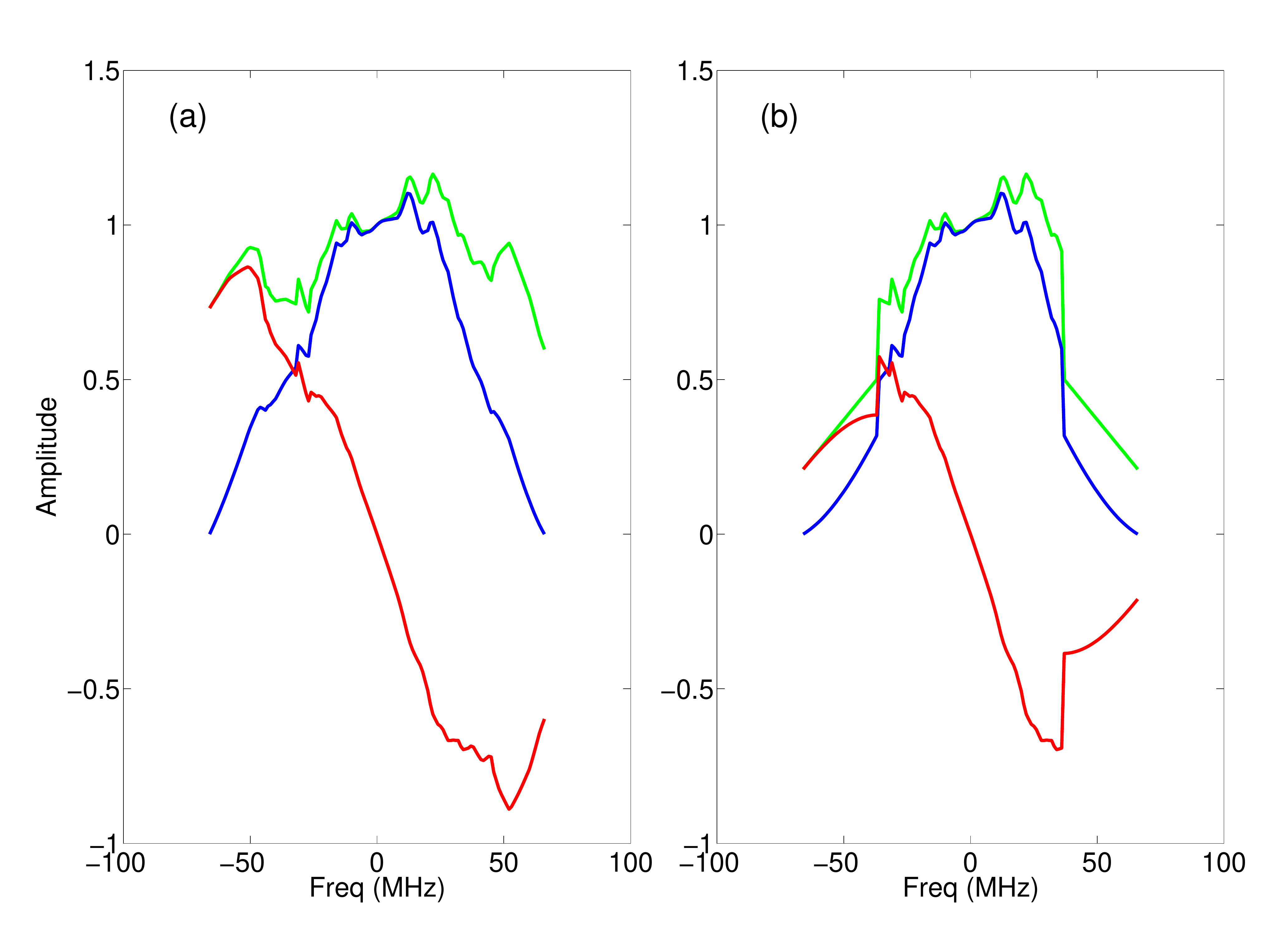}
\caption{(Color online) The transfer function used in open-loop pulse design and simulation. (a) is the experimentally measured transfer function which has the bandwidth of $\sim$130 MHz. (b) is obtained by artificially alteration of (a) to make narrower bandwidth of $\sim$70 MHz. The absolute, real, and imaginary values are in green, blue, and red, respectively. The transfer function is a complex function, and its amplitude and phase are used to adjust the amplitude and phase of our pulses.}
\label{trans}
\end{figure}

Our hardware transfer function $\mathcal{T}$ was measured by observing the Rabi oscillation of the electron spin signal at different offset frequency of the control pulse (see Ref~\cite{PB} for more detail). By fitting the oscillation trace, the response of the spin to the pulse with different frequency can be obtained as shown in Fig.~\ref{trans}(a). This was used in the open-loop control pulse design and simulation. The transfer function in Fig.~\ref{trans}(b) is an artificial alteration from Fig.~\ref{trans}(a) to make the bandwidth narrower for simulation purpose only.

\section{Learning rates in open- and closed-loop methods} 
Both in the open-loop and closed-loop optimization processes, the learning rates ($c_{q}$ in Eq.~(\ref{iterative}) of the main text) affect the performance of the algorithms. We chose the leaning rates in a way that $c\bf g$ have comparable magnitudes for the open-loop and closed-loop optimizations in the first couple of iterations using the same initial pulse shape.  The learning rates were fixed to $c$ when the fidelity was smaller than 0.95, and were $0.5c$ between 0.95 and 0.97, $0.25c$ between 0.97 and 0.98, and finally became $0.125c$ when the fidelity was larger than 0.98. Furthermore, when the fidelity in the experimental closed-loop control got saturated, the learning rates were changed in a larger range to check whether the saturation can be avoided. However, this further optimization of the learning rates did not help a lot. Besides the learning rates, the small difference value $\Delta u^q$ in Eq.~(\ref{finiteG}) of the main text should also be chosen carefully. In the experimental closed-loop control, this value was optimized for the first one or two iterations and then fixed until reaching the fidelity saturation. Near the saturation the value was further tuned to try to break the saturation. However, same as the learning rate, the final tuning of $\Delta u^q$ did not help a lot in obtaining better control fidelities. In the open-loop control, $\Delta u^q$ was chosen to resemble the experimental value and fixed for all the iterations. The optimization codes of the open-loop optimization and simulations for the closed-loop optimization were written using Matlab and can be provided upon request.%(here $P$ is the allowed maximum pulse power)}

\section{The estimated control quality of the open-loop pulses with different transfer functions bandwidths}
Shown in Table~\ref{estma} is the simulated control quality of the open-loop pulses for gate 2. The two pulses considered here were designed using the linear basis set and the Slepian basis set separately, and both of them were designed using the condition that $A{=}$66 MHz, $B{=}$26 MHz, $\mathcal{T}{=}\mathcal{T}_{meas}$. The simulated values show that the accurate transfer function information is important in the open-loop pulse design, which is different from the closed-loop case.
\begin{table}[h]
\normalsize
\centering
\begin{tabular}{C{18mm} C{25mm} C{25mm}}
	\hline\hline
	FWHM &  linear basis & Slepian basis \\\hline
$\sim$130MHz & 0.991 & 0.990 \\
$\sim$70MHz & 0.898 & 0.899 \\
	\hline\hline
\end{tabular}
\caption{Simulated control quality of the open-loop pulses with different transfer functions bandwidth.}
\label{estma}
\end{table}

\section{Propagation of error in measurements to uncertainties in fidelities}
Here we explain how the errors in measurements (i.e. signal-to-noise ratio or SNR) propagate to the uncertainties in fidelities reported in Table~\ref{Exp} of the main text which are the standard deviations of 50 measurements. Taking $F_{\operatorname{XI}}$ given in Eq.~\ref{control_quality} of the main text as an example, \begin{align}
F_{\operatorname{XI}} &= \frac{1}{2} \left( \frac{S_{L}}{\overline{S}_L}+\frac{S_{R}}{\overline{S}_R} \right)
\end{align} The uncertainty in $F_{\operatorname{XI}}$ is given by \begin{align}
\delta F_{\operatorname{XI}} &= \left[\left(\frac{S_L}{\overline{S}_L}\right)^2\left\{\left(\frac{\delta S_L}{S_L}\right)^2+\left(\frac{\delta \overline{S}_L}{\overline{S}_L}\right)^2\right\} \right. \nonumber \\
& \left. + \left(\frac{S_R}{\overline{S}_R}\right)^2\left\{\left(\frac{\delta S_R}{S_R}\right)^2+\left(\frac{\delta \overline{S}_R}{\overline{S}_R}\right)^2\right\} \right]^{1/2}
\end{align} using the error propagation rules for addition (if $Q{=}a{+}b$ then $\delta Q{=}\sqrt{(\delta a)^2{+}(\delta b)^2}$) and multiplication/division (if $Q{=}a/b$ then $\delta Q/|Q|{=}\sqrt{(\delta a/a)^2{+}(\delta b/b)^2}$). $\delta F_{\operatorname{XI}}$, $\delta S_{L}$, $\delta \overline{S}_L$, $\delta S_{R}$, and $\delta \overline{S}_R$ denote the uncertainties in $F_{\operatorname{XI}}$, $S_{L}$, $\overline{S}_L$, $S_{R}$, and $\overline{S}_R$, respectively. For HQCA, $S_L/\overline{S}_L{\sim}1.00$, $\delta S_L/S_L{\sim}0.03$, $\delta \overline{S}_L/\overline{S}_L{\sim}0.02$, $S_R/\overline{S}_R{\sim}0.94$, $\delta S_R/S_R{\sim}0.02$, $\delta \overline{S}_R/\overline{S}_R{\sim}0.02$ which gives $\delta F_{\operatorname{XI}}{\sim}0.04$ which agrees well with the reported standard deviation. The values for other closed-loop methods and open-loop conditions are very similar, and the uncertainty in $F_{\operatorname{ZZ}}$ is mathematically equivalent to that of $F_{\operatorname{XI}}$.

\section{Effect of additional H spins in the environment}
Throughout the paper, we use the simplified, two-spin Hamiltonian with one electron and one $\alpha$-H spin. However, there are more H spins in the environment that are weakly coupled to the electron spin (Ref.~\cite{MAENDOR} identifies seven more H spins with an order of magnitude smaller coupling strengths compared to the $\alpha$-H spin). Such weak couplings are not resolvable in the ESR spectrum because the linewidth is too broad ($\sim$10 MHz; see Fig.~\ref{fig1}(c) of the main text). And we treated the effect of the weak environment H spins as a static inhomogeneous local field. However, these environment H spins will likely change their state during our control pulse. In order to gain some insight about the influence of additional environment H spins on our closed-loop optimal control, further simulation of $\operatorname{ZI-ZZ}$ gate using the FD method was performed with one electron, one $\alpha$-H, and one weakly coupled H spin. For the weakly coupled H spin, different coupling strength constants of $A$ and $B$ were considered. It was found that while the strength of $A$ has little effect, larger $B$ value tends to slow down the process. For example, the simulation with $A{=}$4 MHz and $B{=}$0 MHz generated the same fidelity value compared to the simulation without the weakly coupled H, under same termination condition (\textit{e.g.} when the average fidelity improvement is less than 0.002 per iteration). However, when $A{=}$0 MHz and $B{=}$4 MHz, the final fidelity was 0.004 lower than the simulation without the weakly coupled H. Therefore, the environment H spins can be another error source, especially if they have non-zero $B$. Due to large amount of computing resources required and uncertainty in the crystal orientation, we did not simulate all the seven H spins identified by Ref.~\cite{MAENDOR}. The effect of weakly coupled environment H spins on closed-loop optimal control is of future interest.

%\vspace{12pt}

%\begin{thebibliography}{10}

%\bibitem{ESRRBs} D. K. Park, G. Feng, R. Rahimi, J. Baugh, and R. Laflamme, Journal of Magnetic Resonance 267, 68–78 (2016).

%\bibitem{khaneja2005optimals}
%N.~Khaneja, T.~Reiss, C.~Kehlet, T.~Schulte-Herbr{\"u}ggen, and S.~J. Glaser, Journal of Magnetic Resonance 172~(2),
%  296--305 (2005).
  
%\bibitem{T2star} The $T_2^*$ values are estimated from the measured spectra using a single lorentzian line shape. In our simulations,  we try to more closely match the measured spectra by using %Larmor frequency distributions composed of multiple Lorentzian and/or Gaussian line shapes instead of just one.%we tried to mimic the measured distributions as closely as possible instead of using single lorentzian line shape.  

%\bibitem{GST1s} S. T. Merkel, J. M. Gambetta, J. A. Smolin, S.
%Poletto, A. D. C\'{o}rcoles, B. R. Johnson, C. A. Ryan, and M.
%Steffen,  Phys. Rev. A 87, 062119 (2013).

%\bibitem{GST2s} R. Blume-Kohout, J. K. Gamble, E. Nielsen, J.
%Mizrahi, J. D. Sterk, and P. Maunz, arXiv:1310.4492 (2013).

%\bibitem{Paulitransfer} J. M. Chow \textit{et al.}, Phys. Rev. Lett. 109, 060501 (2012).

%\expandafter\ifx\csname url\endcsname\relax
%  \def\url#1{\texttt{#1}}\fi
  %
%\expandafter\ifx\csname urlprefix\endcsname\relax\def\urlprefix{URL }\fi
%\expandafter\ifx\csname href\endcsname\relax
 % \def\href#1#2{#2} \def\path#1{#1}\fi

%\end{thebibliography}

\end{document}